\documentclass[a4paper, 11pt, oneside]{article}

\usepackage{times}
\usepackage[T1]{fontenc}	%Permite fazer hifenização
\usepackage[utf8]{inputenc}	%Permite digitar acentos no teclado
\usepackage{amsmath, amsfonts, amssymb, dsfont} %Simbolos Matemáticos
\usepackage{mathtools}	% Melhora a aparência de textos com muitas fórmulas e símbolos matemáticos
\usepackage{amsthm}			%Permite usar o \newtheorem
\usepackage[left=3cm, right=3cm, top=2.5cm, bottom=2.5cm]{geometry}
\usepackage{color, graphicx}
\usepackage{indentfirst}	%Coloca margem no primeiro parágrafo
\usepackage[pdftex]{hyperref}	%Coloca hiper-rifem
\usepackage{epigraph}
\usepackage{caption}
\usepackage{xfrac}
\usepackage{pdfpages}
\usepackage{xspace,setspace}

\newcommand{\be}{\begin{equation}}
\newcommand{\ee}{\end{equation}}
\newcommand{\ben}{\begin{eqnarray}}
\newcommand{\een}{\end{eqnarray}}

\begin{document}

\noindent {\fontsize{16pt}{\baselineskip}\selectfont {\textbf{The Eigenfunction Method to Calculate the Klein--Gordon Propa-gator in an Inhomogeneous Magnetic Field}}

\vspace{12pt}

\begin{flushleft}
\noindent {\fontsize{11pt}{\baselineskip}\selectfont E. B. S. Corr\^{e}a*$^1$, M. G. Lima$^1$, S. C. Q. Arruda$^2$}

\end{flushleft}

\singlespace \noindent {\fontsize{9pt}{\baselineskip}\selectfont $^1$Instituto de Ci\^{e}ncias Exatas, Universidade Federal do Sul e Sudeste do Par\'a - UNIFESSPA, Marab\'a, Brazil}

\singlespace \noindent {\fontsize{9pt}{\baselineskip}\selectfont $^2$Faculdade de Ciências Exatas e Tecnologia, Universidade Federal do Par\'a - UFPA, Abaetetuba, Brazil}

\singlespace \noindent {\fontsize{9pt}{\baselineskip}\selectfont E-mail: ecorrea@unifesspa.edu.br}

\vspace{12pt}

\noindent {\fontsize{10pt}{\baselineskip}\selectfont {\bf Abstract:} In this paper, we use the Eigenfunction method (Ritus' method) to calculate the Klein--Gordon propagator under an external magnetic field with an exponential damping factor $b$, in a four-dimensional Euclidean space. We write the propagator of scalar field in terms of plane waves and Laguerre polynomials. Also the eigenvalues associated to differential operator show the quantization in the $\mathrm{xy}$-plane through the Landau levels. We apply the calculated propagator to the mass parameter of an interacting boson system under an exponentially decaying magnetic field.\\

\noindent {\fontsize{10pt}{\baselineskip}\selectfont {\bf Keywords:} Klein--Gordon propagator, Ritus' method, Damped magnetic field.}\\

\noindent {\bf Mathematics Subject Classification:}

\vspace{12pt}

\section{{\fontsize{14pt}{\baselineskip}\selectfont {\bf Introduction}}}
 %  The Glein-Gordon Equation (KGE) arises in the physics in $1927$ as the relativistic generalization approach for Schrodinger's equation in a quantum particle context. Actually, this equation was originally propost by decribe relativistic electrons (today, we khow that electrons are describe by Dirac's equation). The KGE in spide to be Lorentz invariant, showed a probabilistic problem, namely: the probabilistic density was negative. 
   
  % However, the KGE got a important mean in the quantum field theory context. It describe the scalar field. Therefore, quantum particles with zero spin are represents by states that are eigenvectors of scalar quantum field operator, $\hat{\phi}$. Thus, to found out the quantum operator $\hat{\phi}$, we need solve the scalar field equation. 

     Quantum Field Theory (QFT) is the theory of the creation, propagation and annihilation of quantum particles. The probability amplitude of these events happening is given by an important mathematical function, called the Feynman propagator. The propagator is also important to calculate the scattering rate in heavy ion collision processes. According to the spin of the particles, we must take into account different propagators, e.g., particles of zero spin (half) are related to the Klein--Gordon (Dirac) propagator~\cite{Zee}. It is easy to find the propagator of the free scalar field, namely, all we need to do is apply a Fourier transform to the differential equation satisfied by the Klein--Gordon propagator, and then specify the propagator in momenta space.
     
     An interesting subject to explore is QFT under an external electromagnetic field~\cite{Miransky}. This external field can be created in the collision of particle beams in accelerators, such as the LHC. To describe spin zero particles in an external field scenario, we need to couple this external field to the scalar field propagator. This means modifying the Klein-Gordon equation and, consequently, the equation that the propagator satisfies. In this case, the system loses its translational invariance and, for this reason, the Fourier transform is no longer applicable for the calculation of the scalar propagator.
           
    However, in his classic paper of $1951$, J. Schwinger introduced the proper-time method for computing the Feynman propagator under an external field~\cite{Schwinger}. In Schwinger's method, we have a system initially in $D$-dimensions and an additional coordinate with time dimension (the so-called proper- time) is needed to write the propagator under an external field. From a dimensional point of view, in the proper-time approach, we have the change $D \,\rightarrow\, D+1$. The integral equations in proper-time method become difficult to solve, for instance, Schwinger solved the propagator problem under an external field only for constant fields and plane wave fields~\cite{Schwinger,It-Zuber}.
        
     In this perspective, V. Ritus developed an elegant eigenfunction method to calculate the propa-gator under external fields in the 1970s~\cite{Ritus,Ritus1,Ritus2}. The Ritus method consists of solving the eigenvalue equation associated with the specific operator that appears in the field equation modified by the external field. After that, it is proved that the eigenfunctions found form a complete set of eigenfunctions and then it becomes possible to expand the Green's function of the scalar field (Feynman propagator of the scalar field) in terms of these eigenfunctions.
   
    In the $2000$ years, the Ritus method was used to calculate the propagator of a bosonic field with spin $1$ in the context of the electroweak theory~\cite{Elizalde, Elizalde1}. In the following decade, this method was also applied in a fermionic system with spin $1/2$ in low dimensions for the description of graphene~\cite{Mexicanos} and in a Minkowski space-time with four dimensions for the calculation of the chiral condensate~\cite{Iranianos}. We would like to emphasize that when a magnetic field is applied to a charged system, we have the emergence of so-called Landau levels. These energy levels are quantized in the plane orthogonal to the applied magnetic field. For simplicity, it is more common for some authors to make an approximation to deal with Landau levels: they use only the lowest Landau level (LLL).
    
     In what follows, we will consider Ritus' method to calculate the propagator of a spinless field under an  exponentially decaying magnetic field. In other words, let us calculate the Klein-Gordon propagator under an exponentially damped external magnetic field by factor $b$. We have take into account all Landau levels and not only the LLL. In fact, this paper is part of the same line of study that we have done in the Refs.~\cite{EmersonRBEF,EmersonCJP}.
        
     The paper is organized as follows: in Section II, the general aspects of Ritus' method are presented. As we would like to be pedagogical, first we will find the free scalar propagator, that is, without external magnetic field by the eigenfunction method. Next, we apply the Ritus' method to calculate the Klein-Gordon propagator under an exponentially decreasing magnetic field in the $z$ direction. Its found a discrete spectrum in the $xy$-plane for the eigenvalues of the Klein-Gordon operator modified by the external field. In addition, the eigenfunctions are written in terms of complex exponentials and Laguerre functions.  In Section III, we apply the scalar field propagator under an exponentially damped magnetic field calculated in Section II to a system of interacting bosons and we investigate the behaviour of mass parameter in that system over several values of the magnetic field and damping factor $b$. In Section IV, we make some observations and comments on the results. We choose a natural system of units such that $\hbar=c=1$ and a four-dimensional Euclidean space with the four-position vector given by $u \equiv u^{\rho} = (\tau,x,y,z)$.

\section{{\fontsize{14pt}{\baselineskip}\selectfont {\bf Ritus' method}}}

The eigenfunction method is based on the solution of the eigenvalue equation for the operator that comes from the field equation. 

 In general terms, let ${\hat{\mathcal{L}}}$ be the operator that satisfies the equation
\begin{eqnarray*}
{\hat{\mathcal{L}}} \, \mathcal{F} = 0,
\end{eqnarray*}
for a field $\mathcal{F}$.

The Green's function (propagator) associated to ${\hat{\mathcal{L}}}$ satisfies
\begin{eqnarray*}
{\hat{\mathcal{L}}} \, G(u,u^{\prime}) = \pm \, \delta^{4}(u-u^{\prime}).
\end{eqnarray*}

Furthermore, the eigenvalue equation
\begin{eqnarray*}
{\hat{\mathcal{L}}} \, \zeta_{p}(u) = \lambda_{p} \, \zeta_{p}(u),
\end{eqnarray*}
with the eigenvalue $\lambda_{p}$ associated to the operator ${\hat{\mathcal{L}}}$, combined to the complete set formed by eigenfunctions $\zeta_{p}(u)$ given by 
\begin{eqnarray*}
\int \,dp \, \zeta_{p}(u) \, \zeta_{p}^{*}(u^{\,\prime})  = \delta^{4}(u-u^{\prime}),
\end{eqnarray*}
allows the expansion of the Green's function of the operator ${\hat{\mathcal{L}}}$ in terms of eigenfunctions $\zeta_{p}$, namely,
\begin{eqnarray*}
G(u,u^{\prime}) = \int \, dp \, \zeta_{p}(u) \, \tilde{g}(p) \, \zeta_{p}^{*}(u^{\prime}),
\end{eqnarray*}
where $\tilde{g}(p) = \pm 1/\lambda_{p}.$

Below, we will apply the Ritus' method in two cases: free scalar field and scalar field under an external exponentially decaying magnetic field.

\subsection{{\fontsize{12pt}{\baselineskip}\selectfont \textbf{\textit{Klein--Gordon free propagator by eigenfunction method}}}}

Free scalar particles are described by the free Klein-Gordon equation, which in Euclidean space 4D is written as
\begin{eqnarray}
(-\partial^{2} + m^{2} )\phi(u) = 0.
\label{Eq. K-G semcampo}
\end{eqnarray}
We assume Einstein's notation: $\partial^{2}=\partial_{\mu}\partial_{\mu} = \partial_{\tau}^2+\partial_{x}^2+\partial_{y}^2+\partial_{z}^2$ and $\phi(u)$ is the free scalar field. 

As we already mentioned, finding Green's function is fundamental to perform calculations in QFT. In the Euclidean space, the Feynman propagator $ G (u,u^{\prime})$ satisfies the following equation (see Ref.~\cite{Ramond})
\begin{eqnarray}
(\partial^{2} - m^{2} ) G (u,u^{\prime}) = - \delta^{4}(u-u^{\prime}),
\label{Eq. K-G semcampo1}
\end{eqnarray}
where $\delta^4(u-u^{\prime}) = \delta(\tau-\tau^{\prime})\delta(x-x^{\prime})\delta(y-y^{\prime})\delta(z-z^{\prime})$ is the Dirac's delta function in $4$D.
%%%%%%%%%%%%%%%%%%%%%%%%%%%%%

We know that the plane waves in Euclidean space, $ \exp{(i{p}_{\alpha} u_{\alpha})} $, are eigenfunctions of $\partial_{\nu}$ with eigenvalue $i{p}_{\nu}$:
\begin{eqnarray*}
\partial_{\mu} \left[\exp{(i{p}_{\alpha} u_{\alpha})}\right] = ip_{\mu}\left[\exp{(i{p}_{\alpha} u_{\alpha})}\right].
\end{eqnarray*}

Since
\begin{eqnarray}
[\partial^{2},\partial_{\nu}]=0,
\end{eqnarray}
we notice that plane waves are also eigenfunctions of $\partial^{2}$ operator with eigenvalue $-p^2$. Furthermore, the plane waves form a complete set of eigenfunctions
\begin{eqnarray}
\int \frac{ d^{4}p}{(2\pi)^{4}} [\exp{(ip_{\mu} u_{\mu})}][\exp{(ip_{\nu} u_{\nu}^{\prime})}]^{*}  =\delta^{4}(u-u^{\prime})
\label{ondas1}	
\end{eqnarray} 
and
\begin{eqnarray}
\int \frac{ d^{4}u}{(2\pi)^{4}} [\exp{(ip_{\mu} u_{\mu})}][\exp{(ip_{\nu}^{\prime} u_{\nu})}]^{*}  =\delta^{4}(p-p^{\prime}).
\label{ondas2}	
\end{eqnarray} 

Thus, we can expand $ G(u,u^{\prime}) $ in terms of these eigenfunctions
%%%%%%%%%%%%%%%%%%%%%%%%%%%%%%%%%%%%
\begin{eqnarray}
G(u-u^{\prime}) &=& \int \frac{d^{4}p}{(2\pi)^{4}} \, \left[\exp{\left(ip_{\mu} u_{\mu}\right)}\right] \,{g}(p) \,\left[\exp{\left(ip_{\nu} u_{\nu}^{\prime}\right)}\right]^{*},
\label{Glivre}
\end{eqnarray}
where $g (p)$ is the propagator in momentum space. Note that the translational invariance was represented by $ G(u,u^{\prime}) \equiv G(u-u^{\prime})$ in Eq.~(\ref{Glivre}). To find $g(p)$, we apply $ (\partial^2 - m^{2})$ to Eq.~(\ref{Glivre}). After that, we use Eq.~(\ref{Eq. K-G semcampo1}) and we take Eq.~({\ref{ondas1}}) into consideration. The result is
%%%%%%%%%%%%%%%%%%%%
\begin{eqnarray}
{g}(p) = \frac{1}{p^{2}+m^{2}},
\label{gemM}
\end{eqnarray}
where $p^{2} = p^{2}_{\tau}+p^{2}_{x}+p^{2}_{y}+p^{2}_{z}$. 

\subsection{{\fontsize{12pt}{\baselineskip}\selectfont \textbf{\textit{ The Klein--Gordon propagator in an exponentially decreasing external magnetic field by eigenfunction method}}}}

Now, let us calculate the scalar propagator under a damped magnetic field by factor $b$, namely ${\bf{B}} = B_{0}\exp(-bx)\hat{z}$. This external field is obtained by $\nabla \times \vec{A}^{ext} $ and we choose the gauge $A_{\mu}^{ext} = \left[\frac{}{}0,0,(B_{0}/b)\exp(-bx),0\right]$.

In this case, the Klein--Gordon equation modified by the external field is given by
\begin{eqnarray}
(-D^{2} + m^{2} )\Phi(u) = 0,
\label{Eq. K-G semcampo}
\end{eqnarray}
where $D^{2} = D_{\mu}D_{\mu}$, being the operator $D_{\mu} \equiv \partial_{\mu} + ie A_{\mu}^{ext}$ due the minimal coupling in Euclidean space~\cite{lawrie1}.

Under this magnetic background, the propagator satisfies the equation
%%%%%%%%%%%%%%%%%%%%%%%%%%%%%
\begin{eqnarray}
(D^{2} - m^{2})G(u,u^{\prime},A^{ext}) = -\delta^{4}(u-u^{\prime}).
\label{eqdeGreenBosons}
\end{eqnarray}
%%%%%%%%%%%%%%%%%%%%%%%%%%%%%

Notice that the commutator between $D^{2}$ and $\partial _{\nu}$ is non zero, i.e.,
\begin{eqnarray*}
[D^{2},\partial_{\nu}] \neq 0 \Rightarrow \mbox{plane waves are not eigenfuntions of} \, D^{2} \, \mbox{operator}.
\end{eqnarray*}

Thus, we have to find the eigenfunctions, $ E_{p}(u)$, of the operator $ D^{2} $ (the so-called Ritus' eigenfunctions) and to check if they form a complete set. After that, we can expand the propagator $G(u,u^{\prime}, A^{ext}) $ in terms of them.

The first step to find the propagator by Ritus' method is to solve the eigenvalue equation
\begin{eqnarray}
D^{2}E_{p}= - p^2E_{p},
\label{eqautovalorB}
\end{eqnarray}
where the minus sign is just a convention. The operator  $D^{2}$ is given by
\begin{eqnarray}
D^{2} &=& \nabla^{2}_{4D} - 2i A_{2}\,\partial_{y} - e^{2} \, A^{2}_{2} \nonumber \\
&=&  \nabla^{2}_{4D} - 2i\frac{\omega}{b}\left(\exp(-bx)\right) \partial_{y} - \frac{\omega^{2}}{b^{2}} \left(\exp(-2bx)\right),
\label{Pidois}
\end{eqnarray}
where $\nabla^{2}_{4D} = \partial_{\tau}^{2}+\partial_{x}^{2}+\partial_{y}^{2}+\partial_{z}^{2}$ and $\omega \equiv e B_{0}$ is the cyclotron frequency. Notice that $D^{2}$ has two coupling variables $x$ and $y$.

Since 
\begin{eqnarray*}
\left[D^{2},\partial_{\tau}\right] = \left[D^{2},\partial_{y}\right] = \left[D^{2},\partial_{z}\right] = 0,
\end{eqnarray*}
we use the following ansatz to solve Eq.~(\ref{eqautovalorB}) (for some details, see Ref.~\cite{Emerson})
\begin{eqnarray}
{E}_{p}(u) = C \exp\left[i\left(p_{\tau} \tau + \frac{\omega}{b} p_{y} y +p_{z} z\right)\right]X(x),
\label{ansatzeB}
\end{eqnarray}
where $C$ is a normalization constant. The Eq.~(\ref{eqautovalorB}) together with Eq.~(\ref{Pidois}) gives   
\begin{eqnarray}
\left\{\frac{d^2}{dx^2}+ 2 \frac{\omega^2}{b^2}p_{y} \left[\exp(-bx)\right] - \frac{\omega^{2}}{b^{2}}\left[\exp(-2bx)\right]-\frac{\omega^{2}}{b^{2}}p_{y}^{2} - \left(p_{\tau}^{2}+p_{z}^{2}-p^{2}\right)\right\}X(x) = 0. 
\label{eq.HermiteB}
\end{eqnarray}
Let us define the dimensionless variable $\bar{x} = b x$ with $-\infty < \bar{x} < + \infty$. Thus,
\begin{eqnarray*}
\left\{{b^{2}}\frac{d^2}{d{\bar{x}}^2}+ 2 \frac{\omega^2}{b^2}p_{y} \left[\exp(-\bar{x})\right] - \frac{\omega^{2}}{b^{2}}\left[\exp(-2\bar{x})\right]-\frac{\omega^{2}}{b^{2}}p_{y}^{2} - \left(p_{\tau}^{2}+p_{z}^{2}-p^{2}\right)\right\}X(\bar{x}) = 0\,\therefore
\end{eqnarray*}
\begin{eqnarray}
\left\{\frac{d^2}{d{\bar{x}}^2} - V(\bar{x}) + a^{2} \right\}X(\bar{x}) = 0,
\label{eq.HermiteB1}
\end{eqnarray}
where we have defined 
\begin{eqnarray*}
V(\bar{x}) \equiv \lambda^{2} p_{y}^{2} - 2 \lambda^{2} p_{y} \left[\exp(- \bar{x})\right] + \lambda^{2}\left[\exp(-2 \bar{x} )\right],
\end{eqnarray*}
and two new dimensionless parameters: $a^{2} \equiv \left(p^{2}-p_{\tau}^{2}-p_{z}^{2}\right)/\,b^2$ \,and\, $\lambda^{2} \equiv \omega^2/\,b^4.$

To solve Eq.~(\ref{eq.HermiteB1}), let us change the variable $\bar{x}$ by $\xi$, as way done in~\cite{Morse}. We get
\begin{eqnarray*}
\xi = 2 \lambda \exp(-\bar{x})  \, \Rightarrow  \, \frac{d^2}{d{\bar{x}}^2} = \xi^{2} \frac{d^2}{d{\xi}^2} + \xi \frac{d}{d{\xi}},
\end{eqnarray*}
with $ 0 < \xi <  \infty .$
The differential equation (\ref{eq.HermiteB1}) write in new variable $\xi$ becomes
\begin{eqnarray}
\left\{\xi^{2} \frac{d^2}{d{\xi}^2} + \xi \frac{d}{d{\xi}} - \frac{\xi^{2}}{4} + p_{y}\lambda \, \xi + (a^2 - \lambda^{2}  p_{y}^{2})\right\}X(\xi) = 0.
\label{eq.HermiteB2}
\end{eqnarray}
%%%%%%%%%%%%%%%%%%%%%%%%%%%%%%%%%%%%
Taking into account Ref.~\cite{Bateman}, pg~$188$, Eq.~($11$), namely
\begin{eqnarray*}
(xz^{\prime})^{\prime} + \left(n + \frac{\alpha + 1}{2} -\frac{x}{4} - \frac{\alpha^{2}}{4x}\right)z = 0, \, \mbox{with solution} \, z = e^{-x/2} x^{\alpha/2} L_{n}^{\alpha}(x),
\label{eq.HermiteB3}
\end{eqnarray*}
where $L_{n}^{\alpha}(x)$ is the Laguerre polynomial associated with positive integers $n$ and $\alpha$. We have, after comparing the last two differential equations, that
\begin{eqnarray}
\left\lbrace
	\begin{array}{lcl}
		p_{y} \lambda =(2n+\alpha+1)/2 \,  \,\,\,\,\,\,\,\,\, & &  \\
		a^{2} - \lambda^{2} \, p_{y}^{2} = - \alpha^{2}/4\, . \,\,\,\,\,\,\,\,\,\,\, & &
	\end{array}
	\right.
	\label{system}
\end{eqnarray}
Solving system (\ref{system}) we find
\begin{eqnarray}
\left\lbrace
	\begin{array}{lcl}
		\alpha = 2 \lambda p_{y} - 2n -1 \,  \,\,\,\,\,\,\,\,\, & &  \\
		p^{2} \rightarrow p^{2}_{n} = p_{\tau}^{2}+p_{z}^{2}+ \omega \, p_{y} (2 n + 1) - (b^2 /4) (2 n+1)^{2}.\, \,\,\,\,\,\,\,\,\,\,\, & &
	\end{array}
	\right.
	\label{system1}
\end{eqnarray}
We have a restriction on integer $n$: 
\begin{eqnarray*}
n = 0,1,2,\cdots, [N], \,\, \mbox{where} \,\, N = \lambda \, p_{y} - \frac{1}{2},  
\end{eqnarray*}
where the notation $[N]$ means the largest integer less than $N$.

Some authors, for exemple Ref.~\cite{Arfken}, page $843$, Eq.~($13.79$), define the Laguerre functions $\psi^{\alpha}_{n}(\xi)$, that are orthogonal
\begin{eqnarray}
\int_{0}^{\infty} d\xi \, \psi^{\alpha}_{n}(\xi) \, \psi^{\alpha}_{n^{\prime}}(\xi) = \delta_{\,n,\,n^{\,\prime}}
\label{OrtoLaguerre}
\end{eqnarray}
and satisfies the closure relation (see Ref.~\cite{Chineses})
\begin{eqnarray}
\sum_{n=0}^{\infty} \psi^{\alpha}_{n}(\xi) \, \psi^{\alpha}_{n}(\xi^{\prime})  = \delta(\xi - \xi^{\prime}),
\label{ClosureRelation}
\end{eqnarray}
where
\begin{eqnarray}
\psi^{\alpha}_{n}(\xi) = \sqrt{\frac{n!}{(n+\alpha)!}} \, \exp(-\xi/2) \,\xi^{\alpha/2} \, L_{n}^{\alpha}(\xi).
\end{eqnarray}

Therefore, the normalized Ritus eigenfunctions $E_{p}(u)$ are
\begin{eqnarray}
{E}_{p}(u) = \frac{1}{(2\pi)^{3/2}} \, \exp\left[i\left(p_{\tau} \tau + \frac{\omega}{b} p_{y} y +p_{z} z\right)\right] \, \psi^{2 \lambda \, p_{y}- 2n-1}_{n}(\xi),
\label{ansatzeB1}
\end{eqnarray}
with $p = p(p_{\tau},n,p_{y},p_{z})$ and $\xi = (2\omega/\,b^{2}) \cdot e^{-b\,x}$. From Eq.~(\ref{ClosureRelation}), it is easy to show that the completeness relation between the eigenfuctions $E_{p}(u)$ is
\begin{eqnarray}
\int dp  \, E_{p}(u) \,E^{*}_{p}(u^{\prime}) = \delta^{4}(u-u^{\prime})
\label{ortoB}
\end{eqnarray}
and the orthogonality is
\begin{eqnarray}
\int \,du \, E_{p}(u) \,E^{*}_{p^{\,\prime}}(u) = \delta\left(p_{\tau}-p_{\tau}^{\prime}\right)\,\delta\left[\frac{\omega}{b}\left(p_{y}-p_{y}^{\prime}\right)\right]\delta\left(p_{z}-p_{z}^{\prime}\right)\,\delta_{\,n,\,n^{\,\prime}},
\label{ortoB1}
\end{eqnarray}
where $du \, = \, d\tau \, d\xi \, dy \, dz$.

Finally, the propagator can be expanded in terms of Ritus' eigenfuctions as 
\begin{eqnarray}
G(u,u^{\prime},A^{ext}) = \int dp \,E_{p}(u) \,{\mathcal{G}}(p) \,{E}^{*}_{p}(u^{\prime}),
\label{Gmag1}
\end{eqnarray}
where ${\mathcal{G}}(p)$ is the propagator in the momenta space. Applying the operator $\left(D^{2} - m^{2}\right)$ in Eq.~(\ref{Gmag1}) and taking into account Eq.~(\ref{eqdeGreenBosons}), Eq.~(\ref{eqautovalorB}) and Eq.~(\ref{ortoB}) we conclude that
\begin{eqnarray}
{\mathcal{G}}(p) = \frac{1}{p_{\tau}^{2}+p_{z}^{2}+ \omega \, p_{y} (2 n + 1) - (b^2 /4) (2 n+1)^{2}+m^{2}}.
\label{Landau}
\end{eqnarray}

The quantization on $\mathrm{xy}$-plane is represented by the Landau levels, $n = 0,1, \cdots, \left[\left(\frac{\omega \cdot p_{y}}{b^{2}}\right)-\frac{1}{2}\right]$. 

Note that the magnetic field ${\bf{B} = B_{0}} \exp(-bx) \, \hat{z}$, in the limit $b \, \rightarrow \,0$, becomes a constant field along to $z$ direction. In this case, the Eq.~(\ref{Landau}) gives correctly  the propagator in the momenta space to a constant magnetic field (see for instance, Ref.~\cite{EmersonRBEF}, Eq. (18), for $\omega \, p_{y} \equiv {\omega_{0}}$).}

\section{{\fontsize{14pt}{\baselineskip}\selectfont System of bosons under an external magnetic field exponentially damped}}

For investigate corrections on the  mass parameter $m^{2}$ of a bosonic system in $D$ Euclidean dimensions with quantum interaction $(\lambda_{0}/4!)\phi^{4}$, being $\lambda_{0}$ the coupling constant and $\phi$ the bosonic field, we can use the following expression~\cite{LucianoPRD2013}
\begin{eqnarray*}
{M}^{2} = m^{2} + \frac{\lambda_{0}}{2}\int \frac{d^{D}p}{(2\pi)^{D}}\,\mathcal{G}(p),
\end{eqnarray*}
where $\mathcal{G}(p) = {1}/{\left({\bf{p}}^{2}+m^{2}\right)}$ is the propagator in momenta space, without magnetic and temperature effects.

Thermal effects over the system are including by Matsubara prescription~\cite{EmersonPRD2019,EmersonPRD2022}
\begin{eqnarray*}
\int \frac{dp_{\tau}}{2\pi} \,\rightarrow\, T \sum_{n_{\tau} = -\infty}^{+\infty} \,\,\,\,\,;\,\,\,\,\,p_{\tau} \,\rightarrow\, \omega_{n_{\tau}} - i\mu,
\end{eqnarray*}
where $T$ is the temperature of the system, $\mu$ its chemical potential and $ \omega_{n_{\tau}} = \pi T\left(2\,n_{\tau}\right)$ are the Matsubara frequencies of field $\phi$.

Now, let us calculate the thermal correction to mass parameter $m^{2}$ of a bosonic system with quantum interaction $(\lambda_{0}/4!)\Phi^{4}(u)$ and under the presence of an external magnetic field with damping factor $b$ along to the $z$ direction in a four-dimensional Euclidean space. In that case, from Eq.~(\ref{Landau}) we have  
\begin{eqnarray*}
M^{2} = m^{2} + \frac{\lambda_{0}}{2}\int \, \frac{dp_{z}}{(2\pi)}\,\, T\sum_{n_{\tau} = -\infty}^{+\infty} \frac{\tilde{\omega}}{2\pi \, b} \sum_{n = 0}^{\tilde{\omega} /b^{2}}  \,\frac{1}{(\omega_{n_{\tau}} - i\mu)^{2}+p_{z}^{2}+ \tilde{\omega} \, (2 n + 1) - (b^2 /4) (2 n+1)^{2}+m^{2}}.
\end{eqnarray*}

In the Figs.~$1$ and $2$ we show the behaviour of $M$ in the temperature range $0 < T < 0.400$~GeV, being $m = 0.130$~GeV the chosen mass parameter of the model.
\begin{figure}[h]
	\centering
	%\subfloat{
	\includegraphics[scale=1.0]{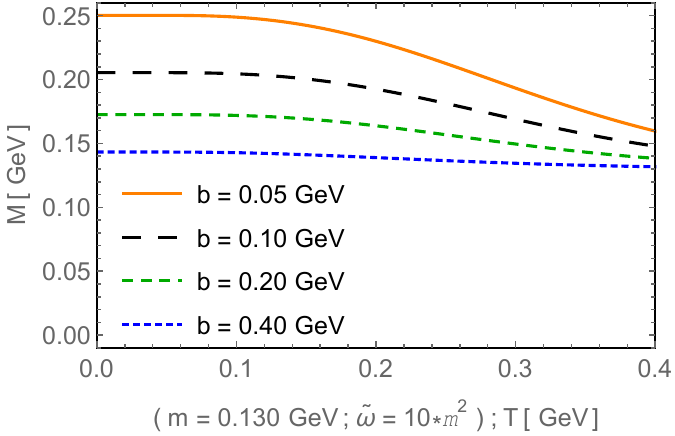}
	\caption{Effective mass $M$ with finite temperature and magnetic effects for several $b$ factors, we fixed $\tilde{\omega} = 10 \cdot m^{2}$ and $\mu = 0$.}
	\end{figure}

\begin{figure}[h]
	\centering
	%\subfloat{
	\includegraphics[scale=1.0]{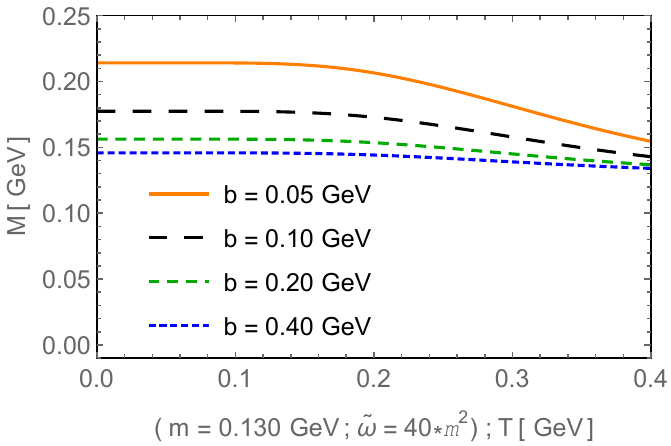}
	\caption{Effective mass $M$ with finite temperature and magnetic effects for several $b$ factors, we fixed $\tilde{\omega} = 40 \cdot m^{2}$ and $\mu = 0$.}
	
\end{figure}

We note that the exponential damping effect of the magnetic field over the system is to make its effective mass smaller as $\tilde{\omega}$ increases. In addition, $b$ factors relatively larger reinforce the decrease in the mass parameter $M$. The system temperature $T$ also makes this parameter smaller. In fact, this last result is well known at constant magnetic fields~\cite{LucianoPRD2013,EmersonPRD2019,EmersonPRD2022}.

\section{\fontsize{14pt}{\baselineskip}\selectfont Conclusion}

In this paper, we apply the Ritus' method to calculate the Feynman propagator of the Klein--Gordon field under an external magnetic field exponentially decreasing and damped by the factor $b$, i.e., we solved the eigenvalue equation for the Klein--Gordon operator modified by the external magnetic field.

After finding the Ritus eigenfunctions, we were able to expand the Green's function of the scalar field in terms of them. We note that the eigenvalue $p^2$ depends on the factor $b$ and $p_{y}$, for an exponentially decreasing external magnetic field. We apply the calculated propagator to a boson system in the temperature range $0 < T < 0.400$~GeV. We found that the system mass parameter $M$ takes on smaller values as $b$ or the magnetic field strength increases.

Ritus originally proposed this method to find the propagator of the Dirac field under an external and constant electromagnetic field. However, the method is very powerful and can be applied in other contexts, as was done here. 

We would like to emphasize that the additional quantum number $k$ present in several papers describing the Ritus' method, for example in~\cite{Mexicanos,Iranianos}, was not necessary in our calculations. In that sense, our base is easier to manipulate than those found in these papers. 

\section*{\fontsize{14pt}{\baselineskip}\selectfont Acknowledgments}

This paper is dedicated to the birth of Emerson B. S. C. Filho.

\section*{\fontsize{14pt}{\baselineskip}\selectfont Conflict of interest statement}

The authors declare that there is no conflict of interest regarding the publication of this paper.


\begin{thebibliography}{99}
	
%\bibitem{Schwinger} J. Schwinger, Phys. Rev. \textbf{82}, 664 (1951).	

%\bibitem{Farina} F. A. Barone, H. Boschi-Filho, C. Farina, Am. J. Phys. \textbf{71}, 483 (2003).
               
 \bibitem{Zee} Zee A. Quantum field theory in a nutshell. 2nd ed. Princeton, N.J: Princeton University Press; 2010. 576 p. 

\bibitem{Miransky}  Miransky VA, Shovkovy IA. Quantum field theory in a magnetic field: From quantum chromodynamics to graphene and Dirac semimetals. Phys Rep. 2015;576:1–209. Available from: doi.org/10.1016/j.physrep.2015.02.003.   

\bibitem{Schwinger} Schwinger J. On gauge invariance and vacuum Polarization. Phys Rev. 1951;82(5):664–679. Available from: doi.org/10.1103/PhysRev.82.664.

\bibitem{It-Zuber} Itzykson C, Zuber JB. Quantum field theory. New York: McGraw-Hill International Book Co; 1980. 705 p. 

\bibitem{Ritus} Ritus VI. Radiative corrections in quantum electrodynamics with intense field and their analytical properties. Ann Phys (N Y). 1972;69(2):555–582. Available from: doi.org/10.1016/0003-4916(72)90191-1.

\bibitem{Ritus1} Ritus VI. Diagonality of the electron mass operators in a constant field. ZhETF Pis Red. 1974;20:135-138. 

\bibitem{Ritus2} Ritus VI. Eigenfunction method and mass operator in the quantum electrodynamics of a constant field. Zh Eksp Teor Fiz. 1978;75:1560-1583.

\bibitem{Elizalde} Elizalde E, Ferrer EJ, de la Incera V. Neutrino self-energy and index of refraction in strong magnetic field: a new approach. Ann Phys (N Y). 2002;295(1):33–49. Available from: doi.org/10.1006/aphy.2001.6203. 

\bibitem{Elizalde1} Elizalde E, Ferrer EJ, de la Incera V. Neutrino propagation in a strongly magnetized medium. Phys Rev D. 2004;70(4):043012. Available from: doi.org/10.1103/PhysRevD.70.043012.

\bibitem{Mexicanos} Murguía G, Raya A, Sánchez Á, Reyes E. The electron propagator in external electromagnetic fields in low dimensions. Am J Phys. 2010;78(7):700–707. Available from: doi.org/10.1119/1.3311656.
 

\bibitem{Iranianos} Sadooghi N, Taghinavaz F. Local electric current correlation function in an exponentially decaying magnetic field. Phys Rev D. 2012;85(12):125035. Available from: doi.org/10.1103/PhysRevD.85.125035. 

\bibitem{EmersonRBEF} Corrêa EBS, Oliveira JE. Método de Ritus para o cálculo do propagador de Feynman. Rev Bras Ens Fis. 2015;37(3):3302-1-3302–6. Available from: doi.org/10.1590/S1806-11173731811. 

\bibitem{EmersonCJP} Corrêa EBS, Bahia CA, Lourenço JA. The bosonic and fermionic propagators under an external electromagnetic field in a Euclidean manifold. Can J Phys. 2022;100(1):1–14. Available from: doi.org/10.1139/cjp-2021-0102.

\bibitem{Ramond} Ramond P. Field theory: a modern primer. 2nd ed., rev. print. Boulder, Colo: Westview Press; 1990. 329 p. 

%\bibitem{Fraga} E. Fraga, T. Kodama, O. Rold\'an, \textit{Anais do XXXIV Encontro Nacional de F\'isica de Part\'iculas e Campos}, Passa Quatro-MG (2013).

%\bibitem{Russos} A.V. Kuznetsov, N.V. Mikheev, G. G. Raffelt, L. A. Vassilevskaya, Phys. Rev. D \textbf{73}, 023001 (2006).


\bibitem{lawrie1} Lawrie ID. Magnetic scaling in superconductors. Phys Rev Lett. 1997;79(1):131–134. Available from: doi.org/10.1103/PhysRevLett.79.131. 

\bibitem{Emerson} Corrêa EBS, Linhares CA, Malbouisson APC. Finite-size, magnetic and chemical-potential effects on first-order phase transitions. Phys Lett A. 2013;377(34–36):1984–1990. Available from: doi.org/10.1016/j.physleta.2013.05.055. 

\bibitem{Morse} Morse PM. Diatomic molecules according to the wave mechanics. II. Vibrational levels. Phys Rev. 1929;34(1):57–64. Available from: doi.org/10.1103/PhysRev.34.57.

\bibitem{Bateman} Bateman H. Higher Transcendental Functions, vol. 2. New York: McGraw-Hill International Book Co; 1953. 410 p.

\bibitem{Arfken} Arfken GB, Weber HJ. Mathematical methods for physicists. 6th ed. Boston: Elsevier; 2005. 1182 p. 

%\bibitem{Chineses} Y. T. Li, R. Wong,  arXiv:1303.1943v1 [math.CA]~(2013). 

\bibitem{Chineses} Li YT, Wong R. Integral and series representations of the dirac delta function. Commun Pure Appl Anal. 2008;7(2):229–247. Available from: doi.org/10.3934/cpaa.2008.7.229.

\bibitem{LucianoPRD2013} Abreu LM, Linhares CA, Malbouisson APC, Malbouisson JMC. Magnetic effects on spontaneous symmetry breaking/restoration in a toroidal topology. Phys Rev D. 2013;88(10):107701. Available from: doi.org/10.1103/PhysRevD.88.107701. 

\bibitem{EmersonPRD2019} Abreu LM, Corrêa EBS, Linhares CA, Malbouisson APC. Finite-volume and magnetic effects on the phase structure of the three-flavor Nambu–Jona-Lasinio model. Phys Rev D. 2019;99(7):076001. Available from: doi.org/10.1103/PhysRevD.99.076001. 

\bibitem{EmersonPRD2022} Abreu LM, Nery ES, Corrêa EBS. Properties of neutral mesons in a hot and magnetized quark matter: size-dependent effects. Phys Rev D. 2022;105(5):056010. Available from: doi.org/10.1103/PhysRevD.105.056010. 


%\bibitem{Iranianos} N. Sadooghi, F. Taghinavaz, arXiv:1203.5634v2 %                                                                                                                                                                                                                                                                                                                                                                                                                                                                                                                                                                                                                                                                                                                                                                                                                                                                                                                                                                                                                                                                                                                                                                                                                                                                                                                                                                                                                                                                                                                                                                                                                                                                                                                                                                                                                                                                                                                                                                                                                                                                                                                                                                                                                                                                                                                                                                                                                                                                                                                                                                                                                                                                                                                                                                                                                                                                                                                                                                                                                                                                                                                                                                                                                                                                                                                                                                                                                                                                                                                                                                                                                                                                                                                                                                                                                                                                                                                                                                                                                                                                                                                                                                                                                                                                                                                                                                                                                                                                                                                                                                                                                                                                                                                                                                                                                                                                                                                                                                                                                                                                                                                                                                                                                                                                                                                                                                                                                                                                                                                                                                                                                                                                                                                                                                                                                                                                                                                                                                                                                                                                                                                                                                                                                                                                                                                                                                                                                                                                                                                                                                                                                                                                                                                                                                                                                                                                                                                                                                                                                                                                                                                                                                                                                                                                                                                                                                                                                                                                                                                                                                                                                                                                                                                                                                                                                                                                                                                                                                                                                                                                                                                                                                                                                                                                                                                                                                                                                                                                                                                                                                                                                                                                                                                                                                                                                                                                                                                                                                                                                                                                                                                                                                                                                                                                                                                                                                                                                                                                                                                                                                                                                                                                                                                                                                                                                                                                                                                                                                                                                                                                                                                                                                                                                                                                                                                                                                                                                                                                                                                                                                                                                                                                                                                                                                                                                                                                                                                                                                                                                                                                                                                                                                                                                                                                                                                                                                                                                                                                                                                                                                                                                                                                                                                                                                                                                                                                                                                                                                                                                                                                                                                                                                                                                                                                                                                                                                                                                                                                                                                                                                                                                                                                                                                                                                                                                                                                                                                                                                                                                                                                                                                                                                                                                                                                                                                                                                                                                                                                                                                                                                                                                                                                                                                                                                                                                                                                                                                                                                                                                                                                                                                                                                                                                                                                                                                                                                                                                                                                                                                                                                                                                                                                                                                                                                                                                                                                                                                                                                                                                                                                                                                                                                                                                                                                                                                                                                                                                                                                                                                                                                                                                                                                                                                                                                                                                                                                                                                                                                                                                                                                                                                                                                                                                                                                                                                                                                                                                                                                                                                                                                                                                                                                                                                                                                                                                                                                                                                                                                                                                                                                                                                                                                                                                                                                                                                                                                                                                                                                                                                                                                                                                                                                                                                                                                                                                                                                                                                                                                                                                                                                                                                                                                                                                                                                                                                                                                                                                                                                                                                                                                                                                                                                                                                                                                                                                                                                                                                                                                                                                                                                                                                                                                                                                                                                                                                                                                                                                                                                                                                                                                                                                                                                                                                                                                                                                                                                                                                                                                                                                                                                                                                                                                                                                                                                                                                                                                                                                                                                                                                                                                                                                                                                                                                                                                                                                                                                                                                                                                                                                                                                                                                                                                                                                                                                                                                                                                                                                                                                                                                                                                                                                                                                                                                                                                                                                                                                                                                                                                                                                                                                                                                                                                                                                                                                                                                                                                                                                                                                                                                                                                                                                                                                                                                                                                                                                                                                                                                                                                                                                                                                                                                                                                                                                                                                                                                                                                                                                                                                                                                                                                                                                                                                                                                                                                                                                                                                                                                                                                                                                                                                                                                                                                                                                                                                                                                                                                                                              [hep-ph]~(2012).

%\bibitem{alemao} A. W\"unsche, J. Phys. A: Math. Gen. \textbf{31}, 8267 (1998).


%\bibitem{Das} A. Das, \textit{Lectures On Quantum Field Theory}, World Scientific Publishing, London (2008).


%\bibitem{Carmem} C. L. R. Braga, \textit{Notas de F\'isica-Matem\'atica}, Editora Livraria da F\'isica, S\~ao Paulo (2006).

\end{thebibliography}
\end{document}